\documentclass[proc]{edpsmath}
\usepackage{graphicx}
\usepackage{dcolumn}
\usepackage{bm}
\usepackage{epsfig}
\usepackage{subfigure}
\usepackage{color}
\usepackage{epstopdf}

\begin{document}
\title{Cooling Effect of the Richtmyer-Meshkov Instability}

\author{F. Mohseni}\address{ETH Z\"urich, Computational Physics for Engineering Materials, Institute for Building Materials, Wolfgang-Pauli-Strasse 27, HIT, CH-8093 Z\"urich (Switzerland) \email{mohsenif@ethz.ch \&\ mmendoza@ethz.ch}}

\author{M. Mendoza}\sameaddress{1}

\author{S. Succi}\address{Istituto per le Applicazioni del Calcolo C.N.R., Via dei Taurini, 19 00185, Rome (Italy) \email{succi@iac.cnr.it}}

\author{H. J. Herrmann}\sameaddress{1}\secondaddress{Departamento de F\'isica, Universidade Federal do Cear\'a, 60455-760 Fortaleza, Cear\'a,
  (Brazil) \email{hjherrmann@ethz.ch}}

\date{\today}
\begin{abstract}
We provide numerical evidence that the Richtmyer-Meshkov (RM) instability contributes to the cooling of a relativistic fluid. Due to the presence of jet particles traveling throughout the medium, shock waves are generated in the form of Mach cones. The interaction of multiple shock waves can trigger the RM instability, and we have found that this process leads to a down-cooling of the relativistic fluid. To confirm the cooling effect of the instability, shock tube Richtmyer-Meshkov instability simulations are performed. Additionally, in order to provide an experimental observable of the RM instability resulting from the Mach cone interaction, we measure the two particle correlation function and highlight the effects of the interaction. 
The simulations have been performed with an improved version of the relativistic lattice Boltzmann model, including  general equations of state and external forces.
 \end{abstract}


\maketitle
\section{Introduction} 

Particles traveling through a compressible fluid generate waves moving at the speed of sound. 
Moreover, if the particles travel faster than the speed of sound of the medium, the disturbances in the fluid are 
confined to the so-called Mach cone. This phenomenon is very common in many natural systems, including astrophysics and high energy physics \cite{falcke1995jet,goncharov1999theory, wilson2004multifluid,PhysRevC.78.034915,xu2008perturbative}, where relativistic fluid effects are important. 
The existence of relativistic shock-waves in the presence of density variations, leads to the appearance of the RM instability, one of the fundamental fluid instabilities, which occurs whenever a shock wave passes through an interface between regions at different densities. This instability was theoretically predicted by Richtmyer \cite{richtmyer1960taylor} and experimentally detected by Meshkov \cite{meshkov1969instability}, in the non-relativistic context.  The study of the RM instability is of major importance in several fields, ranging from high energy physics \cite{goncharov1999theory, wilson2004multifluid,PhysRevC.78.034915,xu2008perturbative} to astrophysics \cite{arnett2000role} and plasma physics \cite{ottaviani}.  Density variations can appear in the relativistic fluid whenever particles travel through the medium, due to the sweeping effect of the shock waves \cite{Satarov200564}, as well as due to external mechanisms. 

In this work, we show numerically that the RM instability may reduce the average temperature of the relativistic fluid. In particular, we investigate the interaction of two relativistic Mach shocks, and show that the RM instability arises due to this interaction (see Fig.~\ref{2jet}). Furthermore, we find that the appearance of this hydrodynamic instability leads to a decrease in the average temperature of the medium. To justify this finding and to single out the effect of the instability, shock tube RM instability simulations are carried out. The effect of initial domain temperature and density ratio on the cooling effect of the instability is also investigated. Since the growth rate of the instability depends explicitly on its form, the study of the instability can provide information on the equation of state (EoS), the same way shock waves can offer insights on the EoS \cite{stoecker2005collective,PhysRevLett.105.222301}. Thus, gaining information about the RM instability may provide a new means of studying the thermodynamic properties of relativistic fluids. Finally, we propose a way to detect the interaction between Mach cones from an experimental observable, namely the two-point correlation function. For the numerical simulations, the recently developed relativistic Lattice Boltzmann (LB) model \cite{PhysRevD.87.083003} is extended to deal with the ideal gas equation of state and external forces.

This paper is organised as follows. In Sec.~\ref{sec1}, we explain the numerical model and the extensions needed to describe a relativistic fluid with an arbitrary equation of state. The results of our numerical simulations of the RM instability are presented in Sec.~\ref{sec2}, and finally, in Sec.~\ref{sec3} we discuss our results and comment on future works.  
 
\section{Numerical model}\label{sec1}

Let us start the description of the numerical model by presenting the conservation equations for relativistic fluid dynamics, namely
\begin{equation}
\partial_\alpha T^{\alpha \beta}=0, \quad \partial_\alpha N^\alpha=0,
\end{equation} 
where the energy-momentum tensor is defined as $T^{\alpha \beta}=(\epsilon+p) U^\alpha U^\beta/c^2 - p \eta^{\alpha\beta} + \pi^{\alpha \beta}$, and the current density as $N^\alpha=n U^\alpha$ \cite{bass2000dynamics}. Here, $n$ is the number density, $p$ the hydrostatic pressure, $\epsilon$ the energy density, $c$ the speed of light, $\pi^{\alpha \beta}$ the shear-stress tensor, and $\eta^{\alpha\beta}$ the Minkowski metric tensor with the signature $(+,-,-,-)$. The macroscopic four-velocity is $(U^\alpha) = (c, \vec{u}) \gamma (u)$, $\vec{u}$ being the three-dimensional velocity and $\gamma(u)=1/\sqrt{1-u^2/c^2}$ the Lorentz's factor. The Einstein summation convention and natural units i.e., $c=k_B=\hbar=1$, are assumed here and throughout this paper. 

All our numerical simulations are performed using the extended version of the relativistic LB model recently proposed in Ref.~\cite{PhysRevD.87.083003}. This method is a numerical approach based on a minimal lattice version of the relativistic Boltzmann equation, which can be solved to find the probability distribution function in phase space \cite{PhysRevD.87.083003, mendoza2010fast}. The equilibrium distribution $f^{\rm eq}$ for the relativistic Boltzmann equation, in the single-relaxation time approximation for the collision operator, is the Maxwell-J\"uttner distribution function \cite{cercignani}. An extension to simulate high velocities was  also proposed in Ref. \cite{PhysRevD.87.083003}, based on the ultra-relativistic equation of state, i.e., $\epsilon=3p$. However, for this case, the equation for the conservation of energy and momentum is not affected by the density field \cite{PhysRevD.87.083003}, and the two conservation equations become decoupled. This effect suppresses the RM instability, where both equations must be coupled. Therefore, we are interested in a more general ideal gas equation of state, of the form \cite{ryu2006equation}:
\begin {equation}
  p=(\Gamma -1)(\epsilon-n),
\end{equation} 
where $\Gamma=c_p/c_v$, with $c_p$ and $c_v$ being the specific heats at constant pressure and volume, respectively. For low temperatures, i.e. $m c^2/k_B T \ll 1$ , $\Gamma=5/3$, while for high temperatures, i.e. $m c^2/k_B T \gg 1$ , $\Gamma=4/3$. In the ultra-relativistic limit, by replacing $\Gamma=4/3$ and considering the condition $n \ll \epsilon$, the ultra-relativistic equation of state is recovered.

In the relativistic lattice Boltzmann methods based on the model of Marle \cite{marle} for the collision operator, the macroscopic variables can be
calculated by solving a system of equations, corresponding to the moments of the equilibrium distribution \cite{PhysRevD.87.083003, mendoza2010fast}. However, for the case of the ideal gas equation of state, this system of equations cannot be solved due to the fact that the first and the second order moments are coupled. This problem can be solved by using the model of Anderson-Witting \cite{Anderson1974466} for the collision operator. According to the model of Anderson-Witting, the relativistic Boltzmann equation has the following form
\begin{equation}
  p^\alpha \partial_\alpha f=-\frac{U_\alpha p^\alpha}{\tau}(f-f^{\rm eq}),
\end{equation}
where $(p^\mu) = (\gamma, \vec{p})$ is the four-momentum, $f$ is the
probability distribution function, and $\tau$ the single relaxation
time.  The model of Anderson-Witting is based on the Landau-Lifshitz
decomposition \cite{cercignani}. Hence, the macroscopic variables can be
calculated using the following relation \cite{landau}
 \begin{equation}
 U_\alpha T^{\alpha \beta}= \epsilon U^\beta.
\end{equation}
In this case, $\epsilon$ and $U^\beta$ are the largest eigenvalue and
corresponding eigenvector of $T^\beta_\alpha$, respectively. These
values can be calculated numerically using the power method. The density, and the pressure would be evaluated subsequently, using the first order moment relation and the equation of state, respectively.

Additionally, in order to equip the model with an ideal gas equation
of states, the discretised distribution function should be also
modified. Hence, the following term should be added to the original
distribution function proposed in Ref. \cite{PhysRevD.87.083003},
\begin{equation}
\label{IGadd}
  I=\frac{3(\Gamma -1)(\epsilon-n)-\epsilon}{(\Gamma -1)(\epsilon -n) +\epsilon}+ \delta_{i0}\frac{52^2(3(\Gamma -1)(\epsilon-n)-\epsilon)}{33 \times 7^2((\Gamma -1)(\epsilon -n) +\epsilon)},
\end{equation}
where $\delta_{i0}$ is the Kronecker delta function. Note that, as expected, in the ultra-relativistic limit, this term goes to zero. For other details of the numerical model and its validation, one can consult the original article~\cite{PhysRevD.87.083003}.

In the presence of disturbance traveling in a relativistic fluid, the energy-momentum conservation can be expressed as $\partial_{\mu}T^{\mu\nu}=S^\nu$, where the source term $S^\nu$ is the energy deposited by the disturbance and can be written in the form \cite{bouras2012transition,jetbetz}:
\begin{equation}
\label{source}
S^\nu=\frac{1}{(\sqrt{2\pi}\sigma)^2}\exp \left\{-\frac{[\vec{x}-\vec{x}_{jet}]^2}{2\sigma^2}\right\}\left (\frac{d E}{d x},\vec{0}\right ) ,
\end{equation}
where the momentum deposition is ignored and  the disturbance is assumed to travel with the velocity of light. Here $\vec{x}_{jet}$ is the location of the disturbance, where $\sigma=0.04$ and $d E/dx=7.5$ are considered. Numerical units are used here and throughout the paper.

Additionally, to include the external force, $S^\mu$, into the relativistic LB model, we need to calculate the discretised forcing term which will be added to the discretized Boltzmann equation. For the relativistic LB scheme in Ref.~\cite{PhysRevD.87.083003}, and assuming the external force as $(S^\mu)=(S^0,\vec{S})$, the discretized forcing term becomes:
\begin{equation}
S_i=\frac{2 \nu^2 c_0^3}{3c_t^3}(1+I) \frac{\delta t}{\delta x} w_i \left(\Gamma e-(\Gamma -1)n\right) \left[\frac{ c_t S^0}{c_0 e} + \frac{\vec{\xi}\cdot \vec{S}}{p}\right],
\end{equation}
where $I$ is defined in Eq.\eqref{IGadd}, $\vec{\xi}$ and $w_i$ are the discretized lattice vectors and weight functions, respectively, and $\nu$, $c_0$ and $c_t$ are lattice constants which can be found in Ref.~\cite{PhysRevD.87.083003}.

\section{Results}\label{sec2}

For the numerical simulation of the interaction between two Mach cones, two domains with $500 \times 500$ and $250\times 250$ cells are considered. All boundaries are taken as free outlets and the ideal gas equation of state is assumed with $\Gamma=4/3$. The initial position of the disturbance moving along the $x$ direction is $(L/3,L/2)$, while the disturbance moving along the $y$ direction starts at $(L/2,5 L/6)$,  $L$ being the length of the domain. We take $\delta t/\delta x=0.15$ and the controlling parameter of the bulk viscosity is $\alpha=0.25$ (see Ref.\cite{PhysRevD.87.083003}). For more details on the numerical technique, see Ref.~\cite{farhang}.

\begin{figure}
\centering 
\subfigure []{\includegraphics[width=0.25\columnwidth]{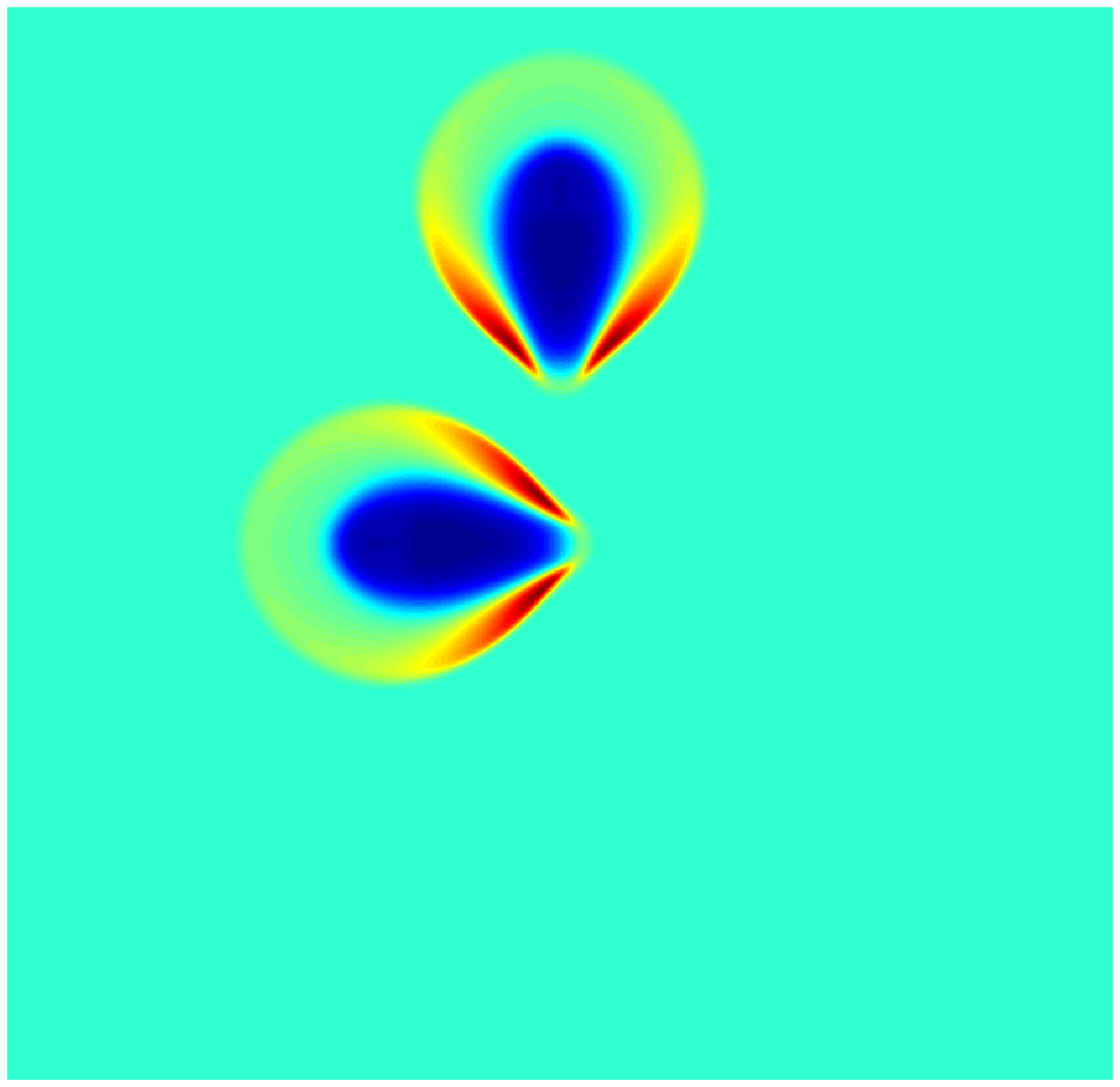} \label{t_3}}
\subfigure []  {\includegraphics[width=0.25\columnwidth]{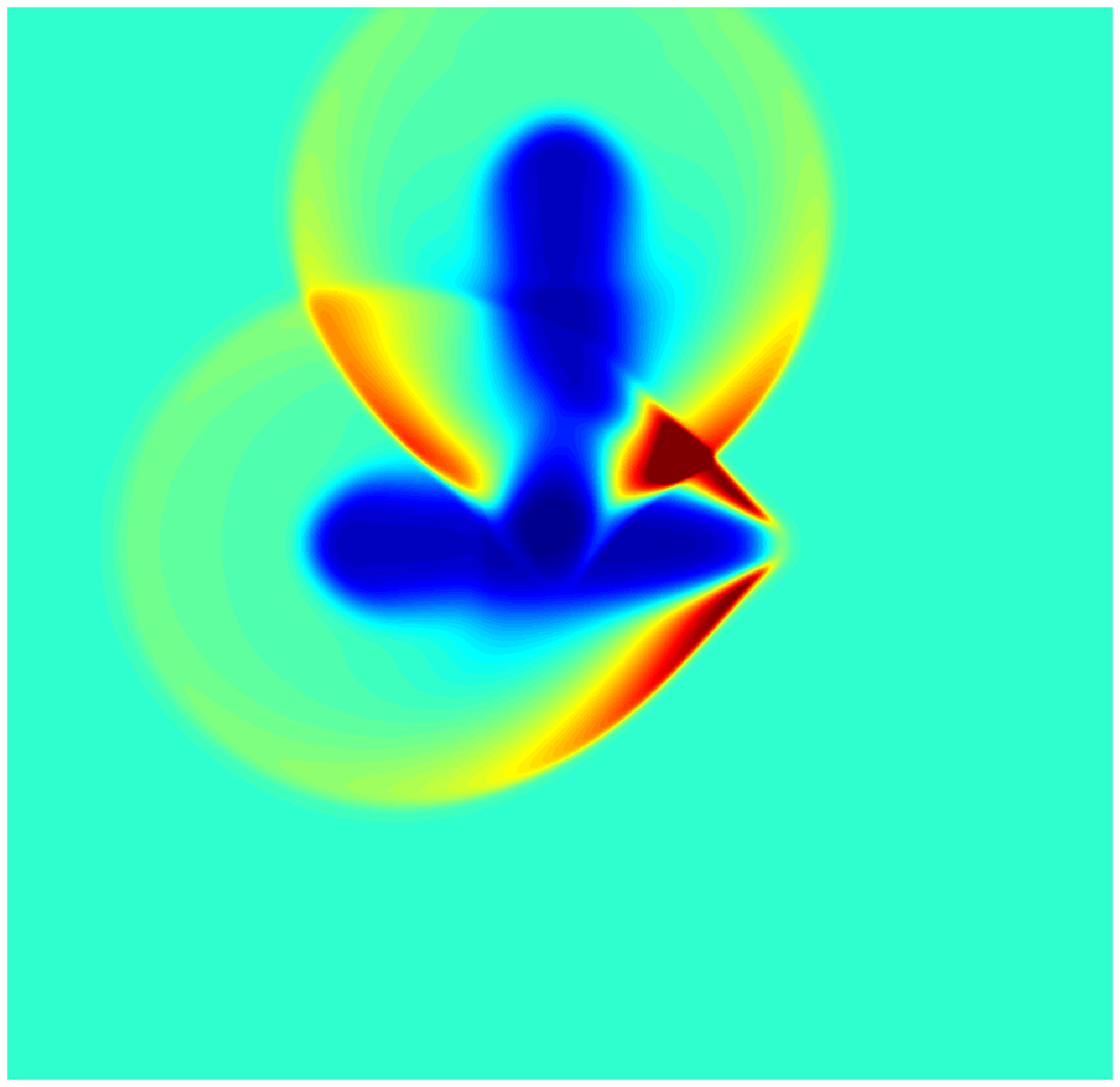}\label{t_6}}
\subfigure []  {\includegraphics[width=0.25\columnwidth]{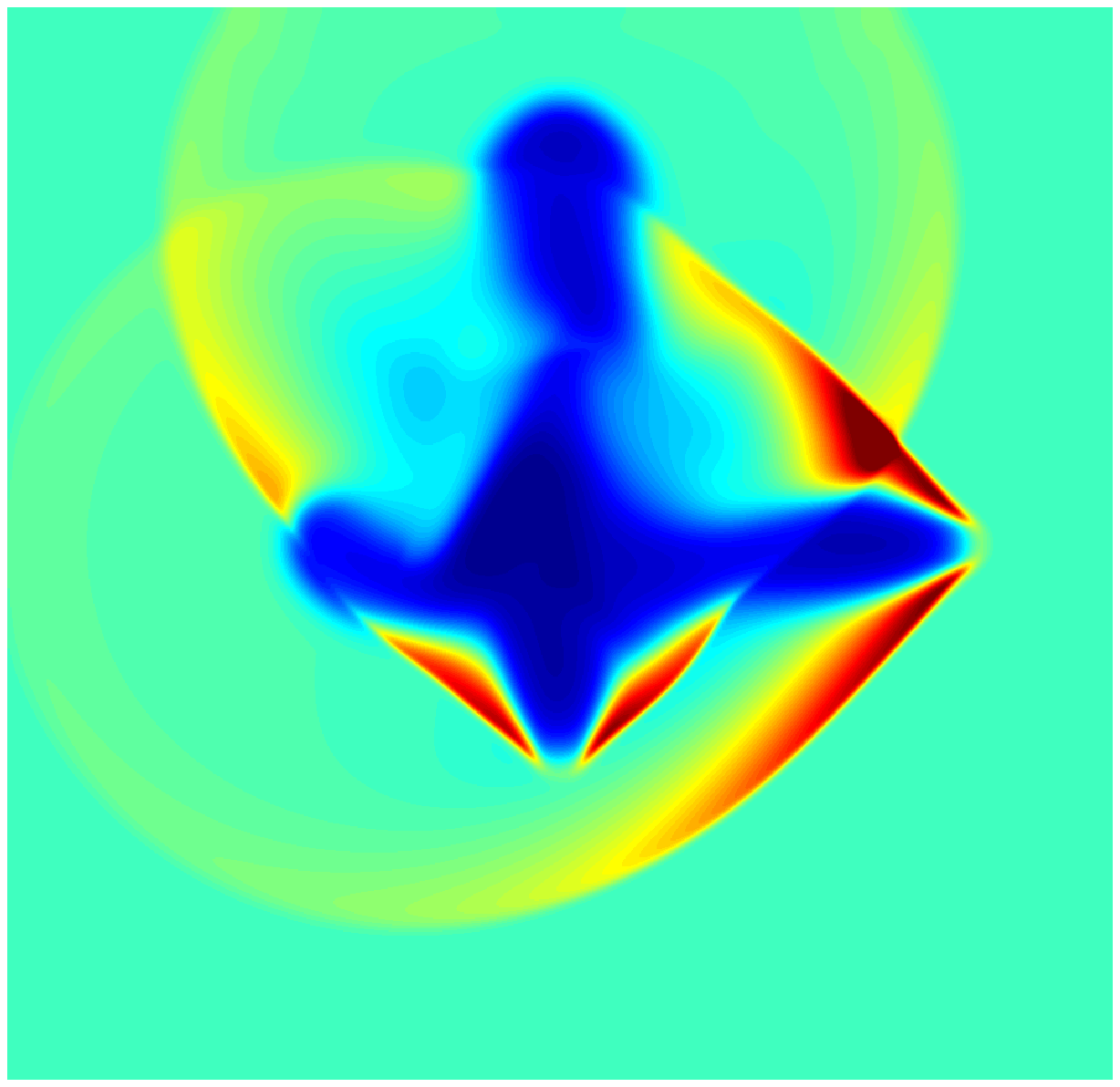}\label{t_9}}
\subfigure []  {\includegraphics[width=0.25\columnwidth]{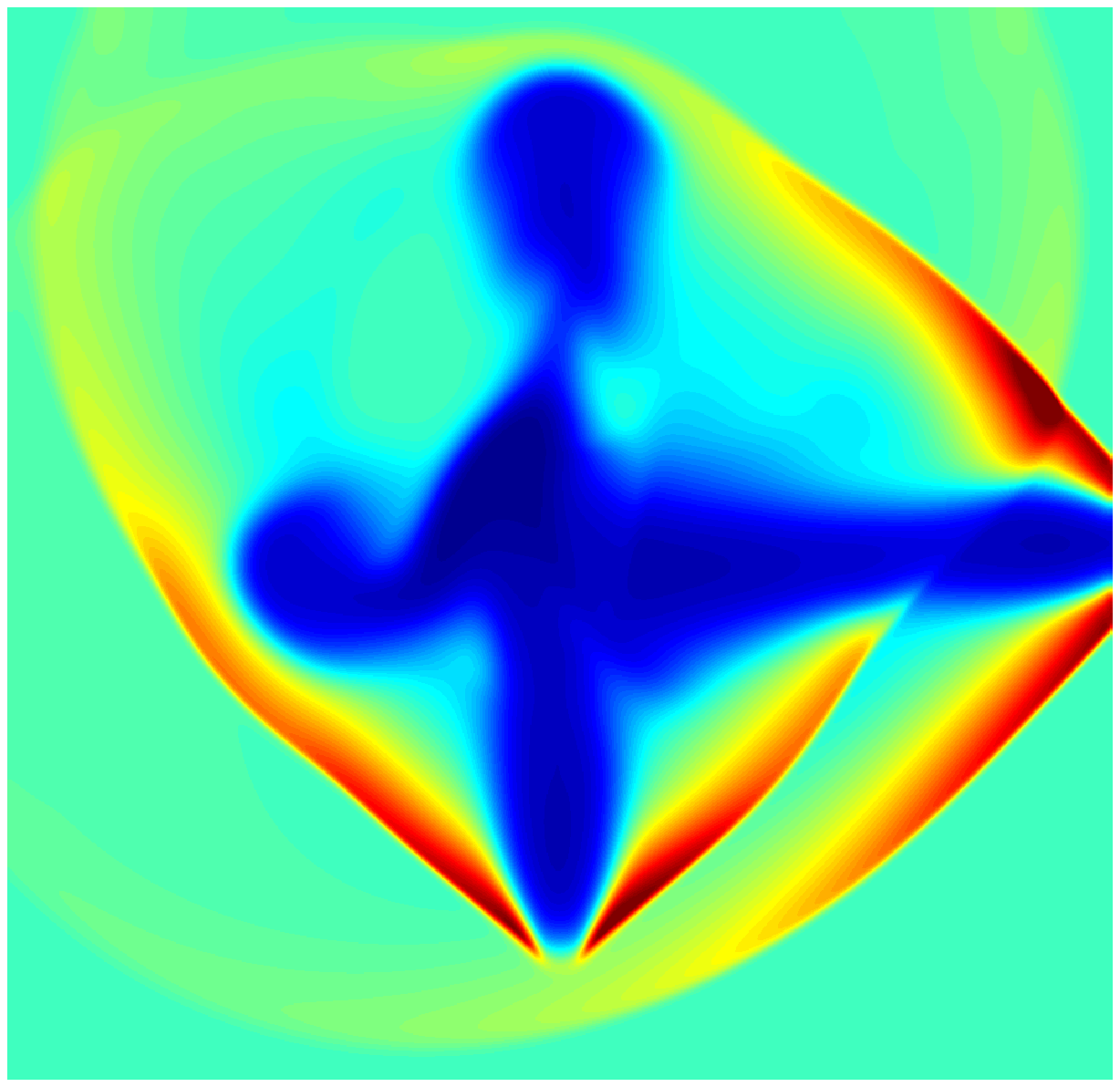}\label{t_12}}
\subfigure []  {\includegraphics[width=0.25\columnwidth]{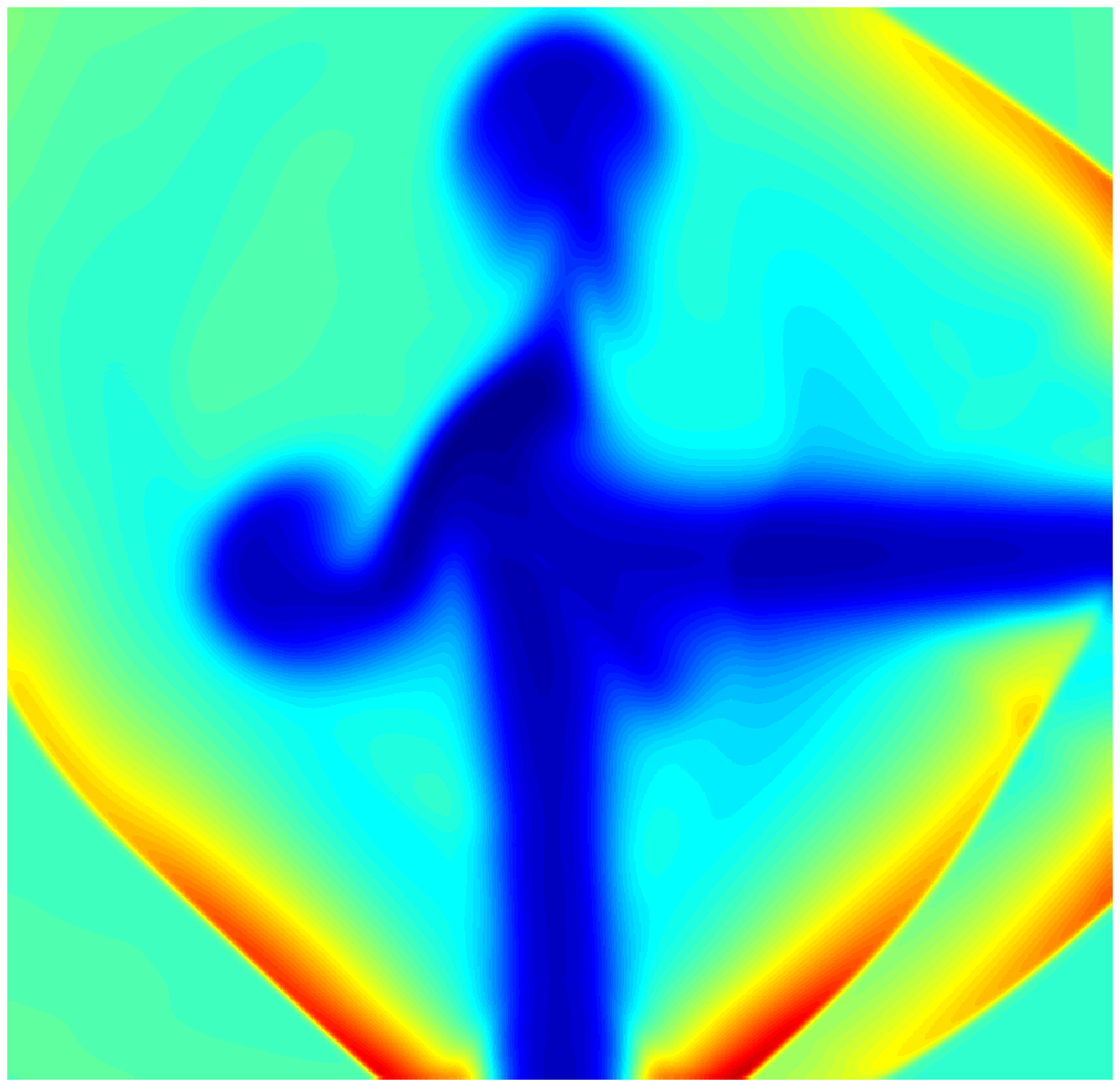}\label{t_16}}
\subfigure []  {\includegraphics[width=0.25\columnwidth]{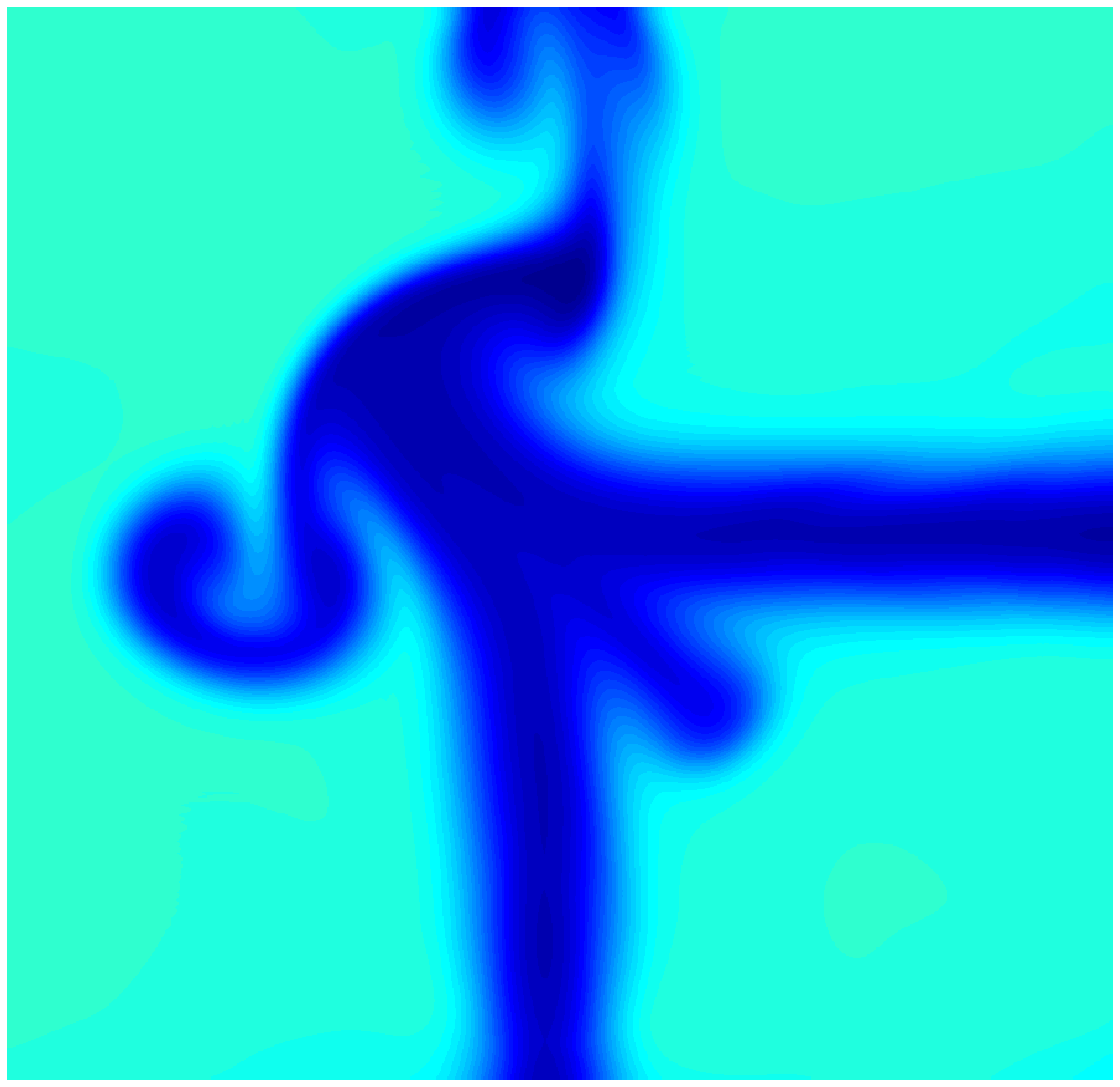}\label{t_20}}
\caption{Snapshots of the density profiles for two interacting Mach cones in a relativistic fluid at times (a) $t=180$ (b) $t=360$ (c) $t=540$ (d) $t=720$ (e) $t=960$, and (f) $t=1200$. Here $T_{med}=0.3$, the red and blue colours denote high and low values of the density, respectively.}
\label{2jet}
\end{figure}

In Fig.~\ref{2jet}, we show that, after the interaction of two Mach cones, when the shock front of a Mach cone passes through the density variation which is caused by the other Mach cone, the RM instability starts to grow in the direction of the advancing shock front. In this figure, we also show that  downstream the moving disturbance and due to the sweeping effect of the shock wave, the density decreases locally. It is worth mentioning that the current simulation shows only one particular case of the interaction of the Mach cones. Nevertheless, what causes the instability is the interaction of one Mach cone with the density fluctuation due to the passage of another Mach cone at earlier times.


\begin{figure}
  \centering 
 \subfigure []{\includegraphics[width=0.4\columnwidth]{2jet_20.eps} \label{t_20_500}}
\subfigure []{\includegraphics[width=0.37\columnwidth]{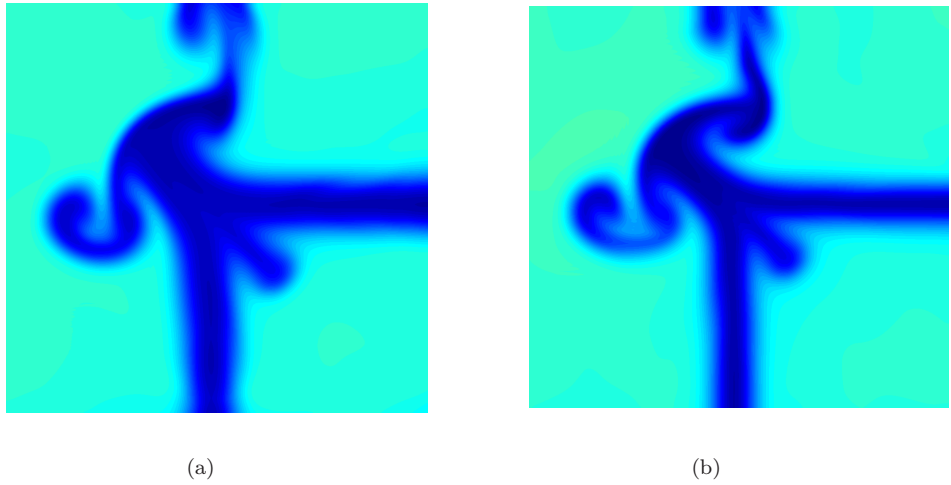}\label{t_20_700}}
\caption{Snapshots of the density profiles for two interacting Mach cones in a relativistic fluid at time $t=1200$ using lattice sizes of (a) $500 \times 500$ and (b) $700 \times 700$ cells. Here $T_{med}=0.3$, the red and blue colours denote high and low values of the density, respectively.}
\label{2jet_comp}
\end{figure}

\begin{figure}
\centering
\includegraphics[width=0.5\columnwidth]{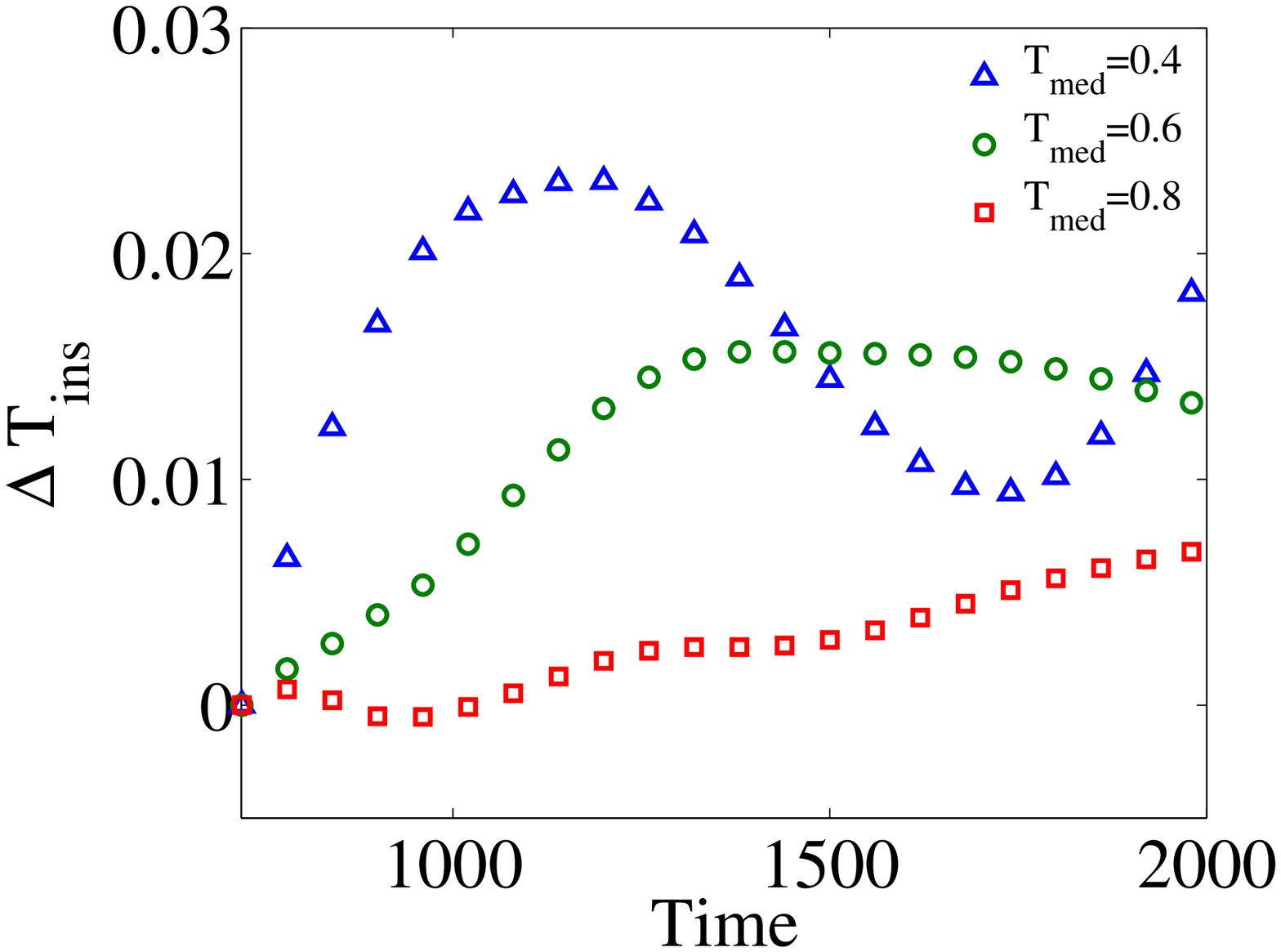}
\caption{Decrease in the average temperature of a relativistic fluid due to the RM instability developed as a result of the interaction of two Mach cones, for different initial temperatures, $T_{med}$. Here $\Delta T_{ins} = T_0 - T'_{med}$, with $T'_{med}$ being the average temperature and $T_0$ the average temperature right after the disturbances have left the simulation zone. Thus, $\Delta T_{ins}$ denotes the decrease in temperature of the media compared with the time when the disturbances and shock waves had left the domain. Here the domain of simulation is $250\times 250$.}.
\label{2jettemp}
\end{figure}

In order to inspect whether the current grid resolution is sufficient to capture the shape of the interaction, additional simulations, with the same parameters as before, but with $700 \times 700$ cells are reported in Fig.~\ref{2jet_comp}. In the same figure, we also show a late stage of the interaction. Despite a slight difference between the results at different resolutions, one can appreciate that the  shape of the instability due to the interaction is well captured with the current $500 \times 500$ lattice resolution.

Regarding the thermal behaviour of the fluid during this phenomenon, as expected, the passing disturbance in the medium increases the average temperature, since it deposits energy to the fluid according to Eq.~\eqref{source}. Here, we are interested in the effects of the aforementioned RM instability on the average temperature. Thus, we compute the average temperature of the medium and compare it to the average temperature when the disturbances and shock waves have left the domain completely. In the absence of the instability, the temperature should remain constant because of the steady state condition. However, we see that, due to the presence of the instability, the temperature starts to decrease, see Fig.~\ref{2jettemp}. The simulation is performed for different initial domain temperatures and we also observe that the decrease in the initial temperature enhances the cooling effect. 
This can be explained by realizing that relativistic effects, which are more dominant at higher temperatures, weaken the RM instability, as recently shown in Ref.~\cite{farhang}. More precisely, the linear growth rate of the instability ($v_f$) is found to take the following form \cite{farhang}:
\begin{equation}
\label{firstpartial}
v_f=\frac{(n_2-n_1)k h_0 \triangle u}{\gamma(2p+\epsilon_2+\epsilon_1)},
\end{equation}
where, $k$ is the initial wave number, $h_0$ is the initial amplitude of the perturbation, $\triangle u$ and $p$ are the velocity and pressure at the interface, respectively, and $\epsilon_2$ ($n_2$) and $\epsilon_1$ ($n_1$) are the energy (density) at both sides of the interface. This expression suggests that the growth rate of the relativistic RM instability decreases as the temperature increases, i.e. the denominator increases in Eq.\eqref{firstpartial}. This is in agreement with the results in Fig.\ref{2jettemp}, where we show that a higher initial temperature leads to a smaller decrease in the average temperature due to the RM instability, since the instability is weaker at higher temperatures.

\begin{figure}
\centering
\includegraphics[width=0.5\columnwidth]{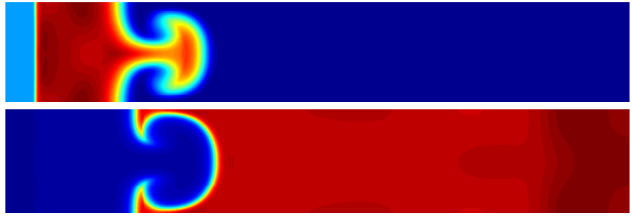}
\caption{Snapshots of the density (top) and temperature (bottom) fields at $t=1260$ in the shock tube RM instability with perturbed interface. Here, we consider the high density ratio $n_L/n_M=28$ and $Ma_r=2.4$. Blue and red colors denote low and high values, respectively. }
\label{contshtu}
\end{figure}

\begin{figure}
\centering
\includegraphics[width=0.5\columnwidth]{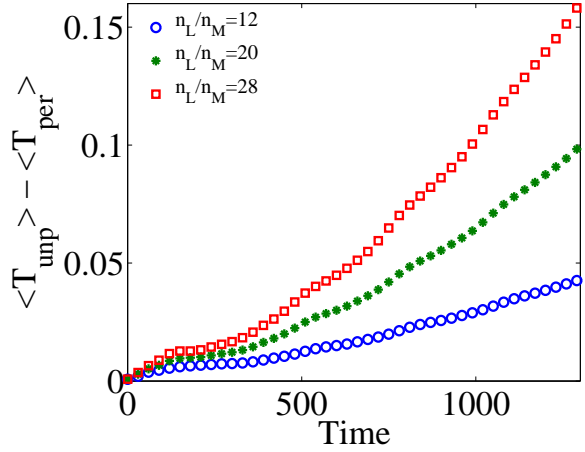}
\caption{Decrease of the average temperature due to the RM instability in a shock tube numerical experiment, for different density ratios. Here $<T_{unp}>$ and $<T_{per}>$ denote the average temperature when the RM instability is (perturbed interface) and is not  (unperturbed interface) present, respectively. We have set $Ma_r = 2.4$. }
\label{densitytime}
\end{figure}

At this point, we have shown that the RM instability can appear during the interaction of Mach cones in a relativistic fluid, and that it contributes to the cooling of the medium. In order to single out the instability and to confirm its cooling effect, we perform simulations in a simplified configuration, namely the shock tube RM instability. The reason for choosing a simple shock tube geometry is twofold; first, this is a standard geometry to study the RM instability, second, it is straightforward to compare the cases with and without the instability. Simulations are performed on a square lattice with $1200 \times 200$ cells. For all simulations considered here, a shock wave with the velocity $\beta = |\vec{u}|/c = 0.94$, travelling from right to left, is passing through a sinusoidal perturbation in the density, located at $x_p = 1000$ cells. The sinusoidal perturbation takes the form, $x_i = x_p + a \sin(\pi/2 + 2\pi y/\lambda)$, where $\lambda$ is the width of the domain and $a = 32$ (for further technical details about the numerical simulation, see \cite{farhang}). Note that subscripts R, M , and L refer to the right hand side of the shock, the region between the shock and the initial perturbation, and the left hand side of the perturbation, respectively. Periodic boundary conditions are considered for the top and bottom boundaries of the domain, while inlet and outlet boundary conditions are applied to the right and left
boundaries, respectively. Other parameters are chosen as follows, $\delta t/\delta x=0.15$, $\alpha=0.25$ and $\Gamma=5/3$. For comparison, the cases without the instability are also simulated by simply setting $a=0$ (unperturbed interface), while other parameters are the same as the ones for the case with the instability. The snapshots of the density and temperature
profiles, for the perturbed case with the density ratio $n_L/n_M=28$ and relativistic Mach number $Ma_r=2.4$ at a late time of the instability, are presented in Fig.~\ref{contshtu} ($Ma_r=u_s \gamma(u_s)/ca \gamma(c_s)$ with $u_s$ being the shock velocity and $c_s$ the sound speed). By measuring the average temperature of the plasma on both cases, with (perturbed interface) and without (unperturbed interface) instability, the decrease in temperature due to the instability can be computed. In Fig.~\ref{densitytime}, one can notice that in the case of a perturbed interface, the average temperature is lower, where the cooling effect of the instability increases in time. 
These numerical experiments are in agreement with the results shown in Fig.~\ref{2jettemp}. The results for different density ratios in Fig.~\ref{densitytime} show that, as the density ratio increases, the decrease in the average temperature is enhanced.  This is because at higher density ratios the instability grows faster, than predicted by Eq.\eqref{firstpartial}. 
\begin{figure}
  \centering 
 \subfigure []{\includegraphics[width=0.5\columnwidth]{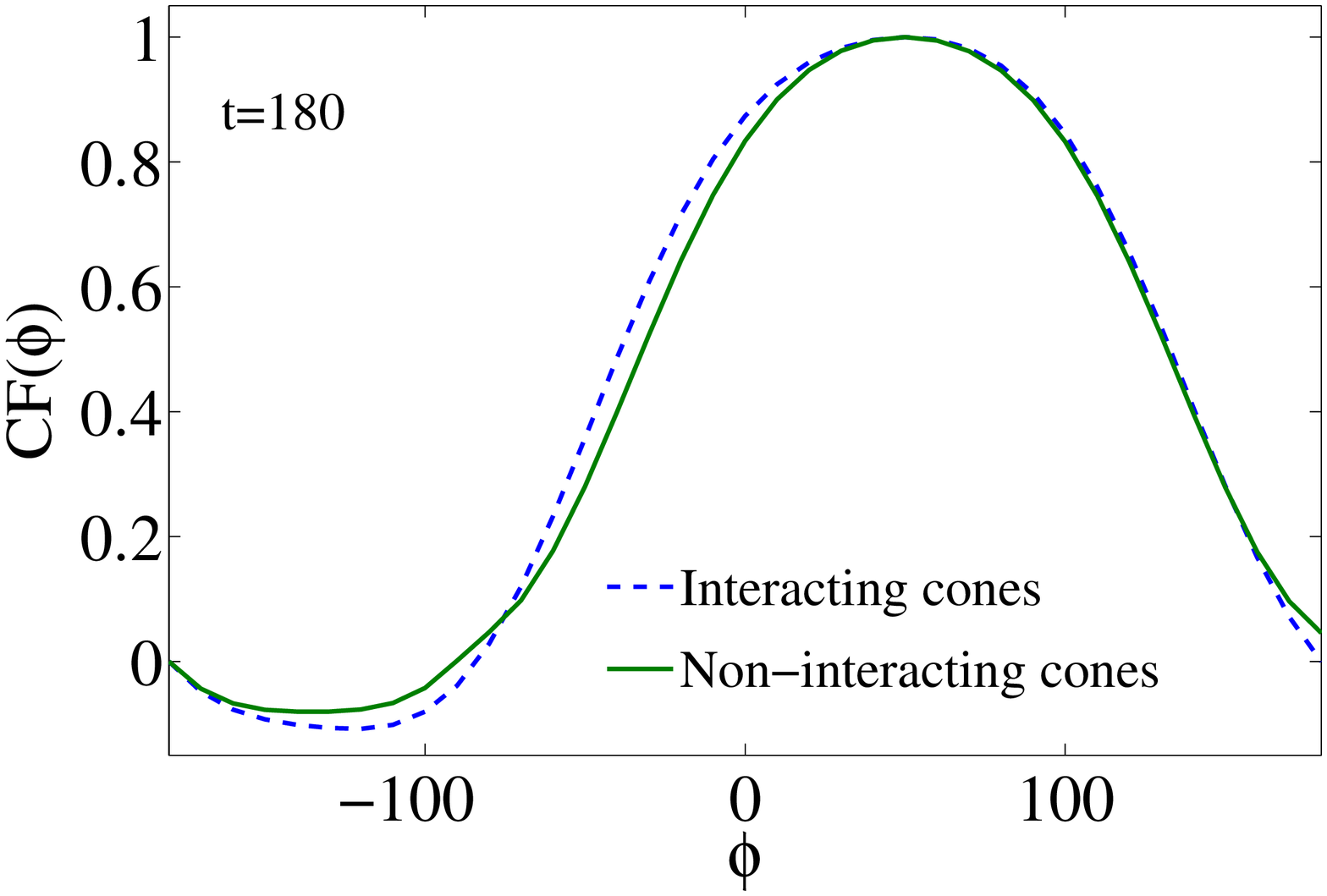}}
\subfigure []  {\includegraphics[width=0.5\columnwidth]{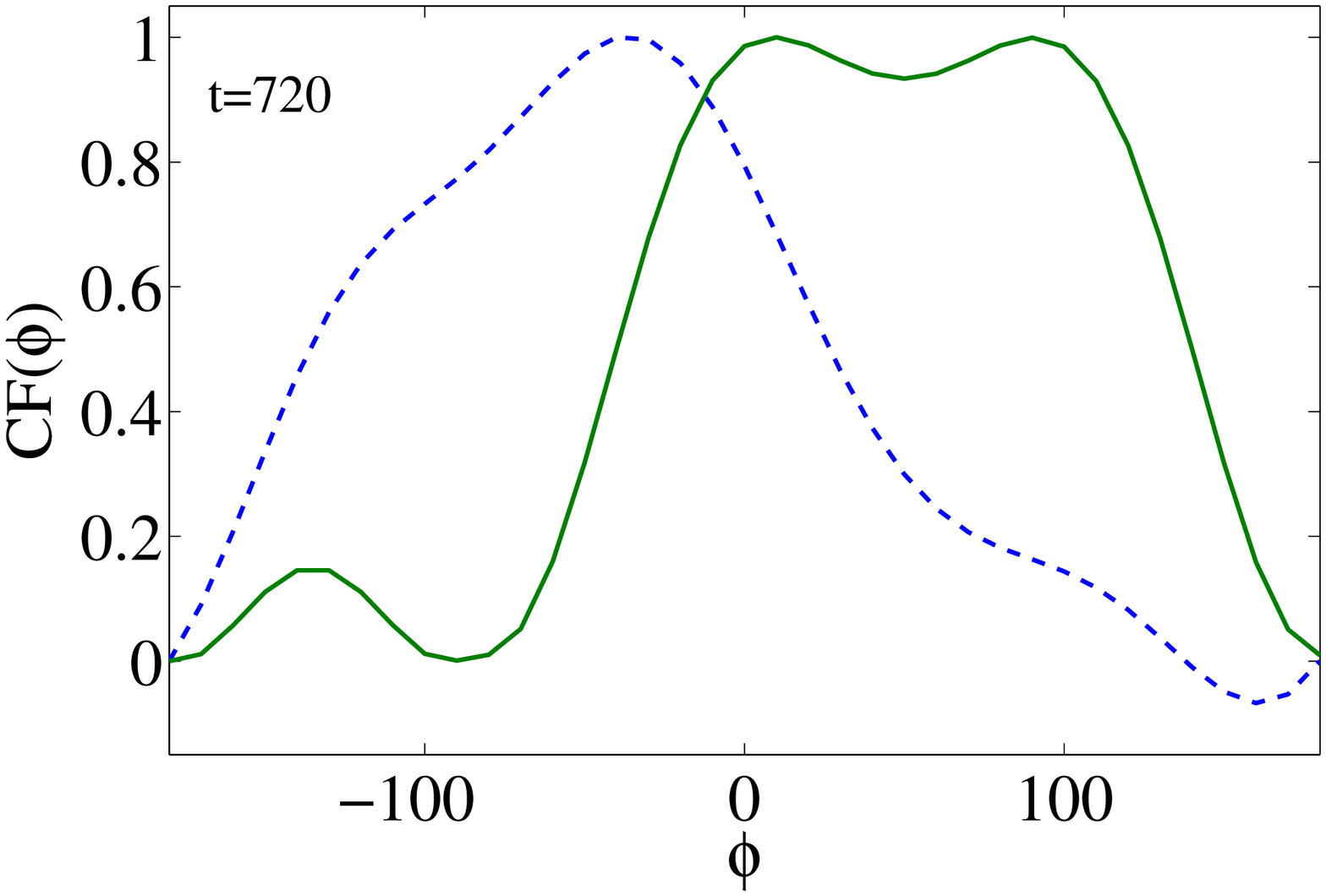}}
  \caption{Two particle correlation (TPC) function for interacting and non-interacting Mach cones for the case $T_{med}=0.3$ in a domain of $500\times500$ at time, (a) $t=180$ (before the interaction), and (b) $t=720$ fm/c (after the interaction).}
\label{2jetcorr}
\end{figure}

Back to the our original problem of Mach cone interaction, we explore the possible experimental observable consequences of the existence of this kind of interactions. To this purpose,  we suggest to study the two-particle correlation (TPC) function \cite{jetbetz,PhysRevLett.105.222301}. Our hydrodynamical calculations provide macroscopic quantities such as temperature and velocity fields. Thus, to compare the hydrodynamical results with experimentally measured observables, a description of the conversion of the fluid into particles is needed. This can be achieved by the Cooper-Frye freeze-out approach \cite{cooper1974single}, where the particle emission pattern is given by:
\begin{equation}
\frac{d N}{p_T d p_T d \phi d y}=\int_{\Sigma}d\Sigma_\mu p^\mu \frac{j}{(2\pi)^2} \exp(-\frac{U^\mu p_\mu}{T}),
\end{equation}
where $p_T=\sqrt{p_x^2+p_y^2}$ is the transversal component of the momentum of a particle$, \phi$ is the azimuthal angle, $j$ is the pre-factor for the Maxwell-J\"uttner distribution function, $y=1/2\ln{\frac{E+p_z}{E-p_z}}$ is the rapidity and $d\Sigma_\mu$ is the integral surface in space-time. 
Using isochronous freeze-out hypersurface, $d\Sigma_\mu=(1,\vec{0})d^2\vec{x}$, we have \cite{jetbetz}:
\begin{eqnarray}
&&\frac{d N}{p_T d p_T d \phi d y}=\nonumber\\&&\frac{j}{(2\pi)^2}\int d^2\vec{x} m_T \cosh{y}\times\exp\{-\frac{\gamma}{T}[m_T\cosh{y}\nonumber\\&&-p_T u_x \cos{\phi}-p_T u_y\sin{\phi}]\}
\end{eqnarray}
where $m_T=E/\cosh{y}$.

Upon defining:
\begin{equation}
\frac{d N_{ass}}{p_T d p_T d \phi d y}=\int_{\Sigma}d\Sigma_\mu p^\mu [\frac{j}{(2\pi)^2} \exp(-\frac{U^\mu p_\mu}{T})-f_0],
\end{equation}
with $f_0$ the Maxwell J\"uttner distribution function at $T=T_{med}$ and $\vec{u}=\vec{0}$, the correlation function at mid-rapidity ($y=0$) is given by \cite{jetbetz}:
\begin{equation}
CF(\phi)=\left.\frac{1}{N_{max}}\left(\frac{d N_{ass}(\phi)}{p_T d p_T d \phi d y}-\frac{d N_{ass}(0)}{p_T d p_T d \phi d y}\right)\right|_{y=0},
\end{equation}
where $N_{max}$  normalizes the correlation. 
In this study the value of $p_T=1$ is considered with lower energy cutoff at $7.6$.

We propose that by measuring the TPC function, one can investigate the existence of the Mach cone interaction. Hence, using our numerical data, the TPC function for two interacting Mach cones is compared to its 
non-interacting counterpart (see Fig.~\ref{2jetcorr}). The TPC function for non-interacting Mach cones is calculated by summing up the TPC function of each Mach cone in the absence of the other one. Fig.~\ref{2jetcorr}a shows that
prior to the interaction, both curves are in good agreement, while after the interaction (see Fig.~\ref{2jetcorr}b), the TPC function for the case of interacting Mach cones differs significantly from the non-interacting case. This shows that if the Mach cones shock waves do not interact with each other, the TPC functions for the non-interacting and interacting cases should remain the same. However, due to the interaction, the functions deviate from each other. 

\section{Conclusion}\label{sec3}

In this paper we have investigated Mach cone interactions and the resulting RM instability in a relativistic fluid. For the numerical simulations, the recently developed LB model for relativistic flows has been extended to handle the ideal gas equation of state and external forces. Our results show that the interaction of two Mach cone shock waves, and in particular the interaction of a shock front of a Mach cone with the density variations generated by the other one, leads to the growth of the RM instability in the direction of the advancing shock front. Regarding the thermal behaviour of this phenomenon, we have shown that the average temperature of the media decreases because of the instability.  

To single out the effect of the instability on this cooling process, we implemented shock tube RM instability simulations, which confirm that the instability causes a decrease in the average temperature. Several simulations have been performed for different initial temperatures and density ratios, which demonstrate that decreasing the initial temperature and/or increasing the density ratio, enhances the cooling. This is in line with the analytical relation for the linear growth rate of the relativistic RM instability, Eq.\eqref{firstpartial}, since decreasing the temperature and/or increasing the density ratio, enhances the instability.   

We have also shown that the interaction of Mach cones significantly affects the TPC. Therefore, comparing the observed TPC with the measured TPC of non-interacting Mach cones, may provide a new tool to experimentally identify the interaction. The results of this paper may be relevant to phenomena characterised by the presence of Mach cones and/or RM instability in astrophysics, high energy physics, and plasma physics.

\begin{acknowledgement}
We acknowledge financial support from the European Research Council (ERC) Advanced Grant 319968-FlowCCS. We are also grateful for the financial support of the Eidgenössische Technische Hochschule Z\"urich (ETHZ) under Grant No. 0611-1. 
\end{acknowledgement}

\bibliographystyle{unsrt}
\bibliography{referencesRM}

\end{document}